# Reversible Data Hiding Based on Two-level HDWT Coefficient Histograms

Xu-Ren Luo[1], Chen-Hui Jerry Lin[2] and Te-Lung Yin[3]

[1] Department of Electrical and Electronic Engineering, Chung Cheng Institute of Technology, National Defense University,
Tahsi, Taoyuan 33509, Taiwan, Republic of China
nrgman.luo@gmail.com

[2] Department of Electrical and Electronic Engineering, Chung Cheng Institute of Technology, National Defense University,
Tahsi, Taoyuan 33509, Taiwan, Republic of China
jerrylin@ndu.edu.tw

[3] Department of Computer Science and Information Engineering, China University of Technology,
Hukou, Hsinchu 303, Taiwan, Republic of China
yintl@cute.edu.tw

## ABSTRACT

*In recent years, reversible data hiding has attracted much more attention than before. Reversibility signifies that the original media can be recovered without any loss from the marked media after extracting the embedded message. This paper presents a new method that adopts two-level wavelet transform and exploits the feature of large wavelet coefficient variance to achieve the goal of high capacity with imperceptibility. Our method differs from those of previous ones in which the wavelet coefficients histogram not gray-level histogram is manipulated. Besides, clever shifting rules are introduced into histogram to avoid the decimal problem in pixel values after recovery to achieve reversibility. With small alteration of the wavelet coefficients in the embedding process, and therefore low visual distortion is obtained in the marked image. In addition, an important feature of our design is that the use of threshold is much different from previous studies. The results indicate that our design is superior to many other state-of-the-art reversible data hiding schemes.*

## 1. INTRODUCTION

Known as lossless, invertible, or distortion-free data hiding, reversible data hiding is a type of delicate and vulnerable technique that is mainly used for quality-sensitive applications such as multimedia content authentication, medical imaging systems, law enforcement, and military imagery. The primary goal in such applications is to recover the original media exactly. The key factors of reversible data hiding include the embedding capacity and visual quality of the marked media, which are for achieving satisfactory performance in various applications[1][2] .
The scheme we present in this paper is an attempt to achieve high-performance reversible data hiding, in which embedding and recovering processes in the frequency domain are devised. The particularities of large wavelet coefficient variance and minor changes in wavelet coefficients following the embedding process are exploited to achieve high capacity and imperceptible embedding. The rest of this paper is organized as follows. In Section 2, previous reversible data hiding schemes and their characteristics will be briefly reviewed in terms of embedding capacity and visual quality. Our proposed scheme is introduced in Section 3. Experimental results and comparative analyses are presented in Section 4. Finally, some conclusions are drawn in Section 5.

## 2. PREVIOUS STUDIES

The related works reported in the literature can be classified into two major types according to the domain for hiding information. In type-I, the schemes work on the transform domain. In 2002, Fridrich et al. [3][4] presented a novel RS scheme starts by dividing the image into three disjoint blocks, and then classifies these blocks into different groups—R, S, and U—using a discriminating function. The flipping function is also introduced to convert an R-group to an S-group and vice versa. The scheme further compresses the R and S groups without loss, and embeds message bits into the state of each group. Although it is a novel technique, the payload is still insufficient for some applications and highly dependent on the compression algorithm. Subsequently, Xuan et al. [5] proposed a new scheme in the integer wavelet transform (IWT) domain. In this scheme, one or multiple middle bit-plan(s) in the high-frequency sub-bands is(are) chosen to embed data bits. In 2003, a difference expansion (DE) scheme was proposed by Tian [6], who applied the integer Haar wavelet transform to an image and exploited the DE technique to embed data bits into the high-frequency coefficients. However, this scheme suffers from the location map problem that it is difficult to achieve capacity control. Alattar [7][8] extended Tian's scheme by generalizing the DE technique to the triplets and quads of adjacent pixels. Kamstra et al. [9][10] improved the DE scheme by predicting the expandable locations in the high-pass band. This scheme improves the efficiency of lossless compression, although the embedding capacity is small. In 2007, Thodi et al. [11] proposed a new scheme combining histogram shifting and prediction-error expansion approaches to remedy the problems of Tian's scheme.

In type-II, most of the studies focus on the spatial domain. Fridrich et al. [12] proposed an invertible method to authenticate digital image using the joint bi-level image experts group (JBIG) lossless compression technique to save free space for data embedding. However, some noisy images may lead to the use of higher bit-planes and result in highly visible distortions. The payload is highly dependent on the lossless compression algorithm. Celik et al. [13][14] employed the generalized least-significant-bit (G-LSB) technique and the context-based adaptive lossless image coding (CALIC) to achieve lossless data hiding. In 2006, Ni et al. [15] utilized multiple pairs of maximum and minimum points of image histogram to achieve reversibility. In the scheme, the pixels are modified for data embedding and extraction. The greater the number of pairs that are selected, the larger the payload becomes. Besides, this scheme can achieve higher embedding capacity through using a multi-level strategy. However, such a strategy may lead to significant overhead and insufficient visual quality. Ni's scheme was also extended by others using a location map rather than the knowledge of the pairs to achieve reversibility [16][17]. Although the scheme is simple comparing to other schemes, significant overhead and insufficient visual quality are critical problems. In 2009, Kim et al. [18] utilized the feature of high spatial correlation between neighboring pixels to achieve high-performance data hiding. The embedding capacity in the paper ranges from 6 to 210 k.

For all of the schemes above, additional overhead is the most difficult problems in the restore process. This paper proposes a new histogram shifting technique in the frequency domain and uses the actual payload to achieve high-performance lossless data hiding.

## 3. THE PROPOSED SCHEME

This section present a new scheme that combines the two-level Haar discrete wavelet transform (HDWT) algorithm and new histogram shifting rules. The HDWT algorithm can produce a complete and lossless reconstruction during synthesis with appropriate filters. Moreover, the histogram shifting rules used in the frequency domain can provide higher flexibility in terms of scalability in resolution and distortion.

In our scheme, an original image is first transformed into a frequency domain and sub-bands in the middle- and high-frequency ranges are then used to create sub-band differences. Each histogram of these sub-band differences is then shifted according to a selected threshold. Message bits can then be hidden in the empty bins of the shifted histograms. Finally, the marked image is reconstructed with the sub-bands carrying and non-carrying hidden message by performing the inverse HDWT algorithm to complete the embedding process. In the extracting process, the corresponding inverse operations can exactly extract the hidden message and recover the original image.

### 3.1. Segmentation algorithm

The two-level HDWT algorithm utilizes the four-band sub-band coding system to decompose an image into a set of different frequency sub-bands. As illustrated in Figures 1 and 2, the size of each sub-band is one eithth of the original image in the spatial domain. The eight different sub-bands can be classified into the low-, middle-, and high-frequency sub-bands. Since the low-frequency sub-band of an image incorporates more energy than the other sub-bands, its coefficients are the most fragile that if any of them are manipulated, a suspect can visibly detect the changes on the spatial domain image. However, if the coefficients in the middle- and/or high-frequency sub-bands are altered, the changes in the spatial domain image will be imperceptible to human eyes. Consequently, this characteristic is generally exploited to hide secret message.

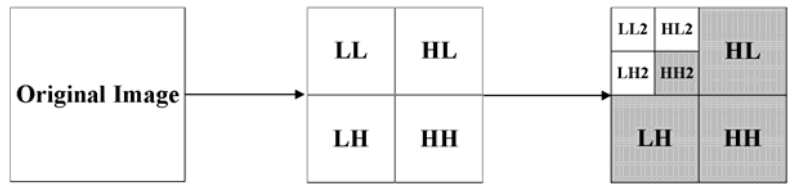

Figure 1. The decomposition process of two-level HDWT

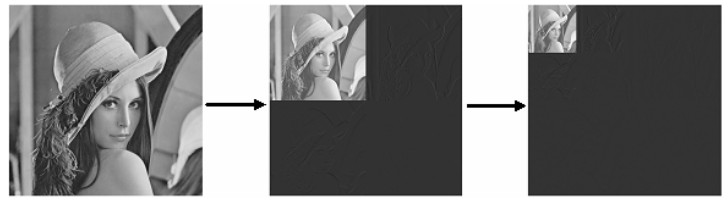

Figure 2. A tow-level HDWT four-band split of "Lena"

### 3.2. Data embedding algorithm

We assume that the secret message is a random binary sequence of 0s and 1s. The histograms of the sub-band differences between the reference sub-band and the other destination sub-bands are shifted to embed the secret message. Figure 3 depicts the overall data embedding process, which is described in detail below.

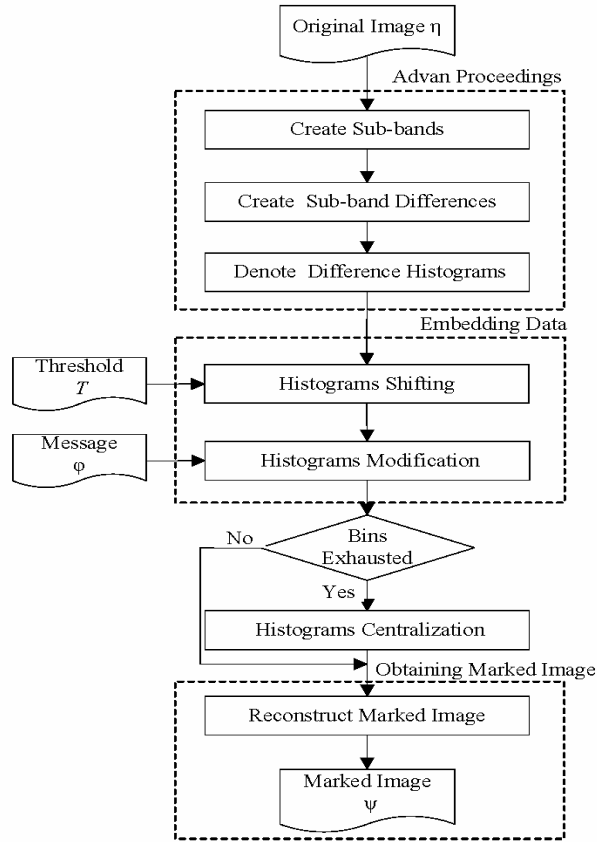

Figure 3. Flowchart of data embedding process

**Embedding_process**$(\eta, T, \varphi)$

Input: $\eta$, the original image; $T$, the threshold according to which the empty bins in each histogram are prepared; $\phi$, the secret message.

Output: $\psi$, the marked image, $f$, the mark of embedding status.

**Advan Proceedings:**

Step 1: Create sub-bands by performing two-level HDWT four-band sub-band coding system on an input image $\eta$. Five of the sub-bands to be utilized are denoted by $LH(x,y)$, $HL(x,y)$, $HH(x,y)$, $LH2(x,y)$, and $HH2(x,y)$, where $(x,y)$ indicates the coordinate of the coefficients in each sub-band.

Step 2: Create sub-band differences $D_1$, $D_2$ and $D_3$ between the reference sub-bands $LH$, $LH2$ and the other destination sub-bands $HL$, $HH$ and $HH2$ by the following formulas:

$$D_1(x,y) = LH(x,y) - HL(x,y), \quad (1)$$

$$D_2(x,y) = LH(x,y) - HH(x,y), \quad (2)$$

$$D_3(x,y) = LH2(x,y) - HH2(x,y). \quad (3)$$

Step 3: Denote the histograms of $D_i$ as $h_i$, where $i = 1, 2, 3$.

Step 4: Shift histogram $h_i$ according to the threshold $T$ selected. The shifted $h_i$ can be calculated as follows:

$$h_i' = \begin{cases} h_i(j) + 8, & \text{if } h_i(j) \geq T + 1, \\ h_i(j) - 8, & \text{if } h_i(j) \leq -(T+1), \end{cases} \quad (4)$$

where $i = 1, 2, 3$ and $(j)$ indicates the value of each bin.

These can also be obtained by the following formulas:

$$D_1'(x, y) = LH(x, y) - HL'(x, y), \quad (5)$$

$$D_2'(x, y) = LH(x, y) - HH'(x, y), \quad (6)$$

$$D_3'(x, y) = LH2(x, y) - HH2'(x, y). \quad (7)$$

where $HL(x,y)$, $HH(x,y)$ and $HH_2(x,y)$ can be expressed as:

$$HL'(x, y) = \begin{cases} HL(x,y) - 8, & \text{if } h_1(j) \geq T + 1, \\ HL(x,y) + 8, & \text{if } h_1(j) \leq -(T+1). \end{cases} \quad (8)$$

$$HH'(x, y) = \begin{cases} HH(x,y) - 8, & \text{if } h_2(j) \geq T + 1, \\ HH(x,y) + 8, & \text{if } h_2(j) \leq -(T+1). \end{cases} \quad (9)$$

$$HH2'(x, y) = \begin{cases} HH2(x,y) - 8, & \text{if } h_3(j) \geq T + 1, \\ HH2(x,y) + 8, & \text{if } h_3(j) \leq -(T+1). \end{cases} \quad (10)$$

**Embedding Data into *HL*, *HH* and *HH2*:**

Step 5: We first set an interation index $\ell$ to T and then embed message bits sequentially by modifying $h_i'$. Each $h_i'$ is shifted to become $h_i''$, where $i = 1, 2$ and $3$ by the following rules:

$$h_i'' = \begin{cases} h_i'(j) + 8, & \text{if } h_i'(j) = \ell, \ \varphi(n) = 0, \\ h_i'(j) + 4, & \text{if } h_i'(j) = \ell, \ \varphi(n) = 1, \\ h_i'(j) - 8, & \text{if } h_i'(j) = -\ell, \ \varphi(n) = 0, \\ h_i'(j) - 4, & \text{if } h_i'(j) = -\ell, \ \varphi(n) = 1, \end{cases} \quad (11)$$

for $\ell > 0$.

$$h_i'' = \begin{cases} h_i'(j) + 8, & \text{if } h_i'(j) = 0, \ \phi(n) = 0, \\ h_i'(j) + 4, & \text{if } h_i'(j) = 0, \ \phi(n) = 1, \end{cases} \quad (12)$$

for $\ell = 0$.

The changes in the difference histograms above result in changes in the coefficients. This implies that the modified sub-band difference $D_i'(x, y)$ is scanned and modified again. Once the value of $D_i'(x, y)$ is equal to $\pm \ell$, the message bit is embedded. This procedure is repeated until there are no $D_i'(x, y)$ with the value of $\pm \ell$. We then decrease $\ell$ by 1 and repeat the step until $\ell < 0$. These steps can be formulated as follows:

$$D_1''(x, y) = LH(x, y) - HL''(x, y) , \quad (13)$$

$$D_2''(x, y) = LH(x, y) - HH''(x, y) , \quad (14)$$

$$D_3''(x, y) = LH2(x, y) - HH2''(x, y) . \quad (15)$$

if $\ell > 0$,

$$HL''(x,y) = \begin{cases} HL'(x,y)+4, & \text{if } h_1'(j) = -\ell, \phi(n) = 1, \\ HL'(x,y)-4, & \text{if } h_1'(j) = \ell, \phi(n) = 1, \\ HL'(x,y)+8, & \text{if } h_1'(j) = -\ell, \phi(n) = 0, \\ HL'(x,y)-8, & \text{if } h_1'(j) = \ell, \phi(n) = 0. \end{cases} \quad (16)$$

$$HH''(x,y) = \begin{cases} HH'(x,y)+4, & \text{if } h_2'(j) = -\ell, \phi(n) = 1, \\ HH'(x,y)-4, & \text{if } h_2'(j) = \ell, \phi(n) = 1, \\ HH'(x,y)+8, & \text{if } h_2'(j) = -\ell, \phi(n) = 0, \\ HH'(x,y)-8, & \text{if } h_2'(j) = \ell, \phi(n) = 0. \end{cases} \quad (17)$$

$$HH2''(x,y) = \begin{cases} HH2'(x,y)+4, & \text{if } h_3'(j) = -\ell, \phi(n) = 1, \\ HH2'(x,y)-4, & \text{if } h_3'(j) = \ell, \phi(n) = 1, \\ HH2'(x,y)+8, & \text{if } h_3'(j) = -\ell, \phi(n) = 0, \\ HH2'(x,y)-8, & \text{if } h_3'(j) = \ell, \phi(n) = 0. \end{cases} \quad (18)$$

if $\ell = 0$,

$$HL''(x,y) = \begin{cases} HL'(x,y)-4, & \text{if } h_1'(j)=0, \phi(n)=1, \\ HL'(x,y)-8, & \text{if } h_1'(j)=0, \phi(n)=0. \end{cases} \quad (19)$$

$$HH''(x,y) = \begin{cases} HH'(x,y)-4, & \text{if } h_2(j)=0, \phi(n)=1, \\ HH'(x,y)-8, & \text{if } h_2(j)=0, \phi(n)=0. \end{cases} \quad (20)$$

$$HH2''(x,y) = \begin{cases} HH2'(x,y)-4, & \text{if } h_3(j)=0, \phi(n)=1, \\ HH2'(x,y)-8, & \text{if } h_3(j)=0, \phi(n)=0. \end{cases} \quad (21)$$

**Centralizing Histogram**

Step 6: When all the bins in the difference histogram are exhausted to hide data, eight bins, valued from –4 to 3, will become empty. In this case, the mark $f$ is set to be "1" and all bins on the right side will be moved left 4 bins, and those on the left will be moved right 4 bins in order to improve the image quality by decreasing the variance of the differences in the coefficients. Otherwise, the mark $f$ is set to be "0". Each $h_i''$ is shifted to become $h_i'''$ by the following rules:

$$h_i'''(j) = \begin{cases} h_i''(j) - 4, & \text{if } h_i''(j) > 0, \\ h_i''(j) + 4, & \text{if } h_i''(j) < 0, \end{cases} \quad (22)$$

where $i = 1, 2$ and $3$.

Shifting histograms as above creates coefficient changes, which can be formulated as follows:

$$D_1'''(x, y) = LH(x, y) - HL'''(x, y), \qquad (23)$$

$$D_2'''(x, y) = LH(x, y) - HH'''(x, y), \qquad (24)$$

$$D_3'''(x, y) = LH2(x, y) - HH2'''(x, y), \qquad (25)$$

where $HL'''(x, y)$, $HH'''(x, y)$ and $HH_2'''(x, y)$ can be expressed as:

$$HL'''(x, y) = \begin{cases} HL''(x, y) + 4, & \text{if } h_1''(j) > 0, \\ HL''(x, y) - 4, & \text{if } h_1''(j) < 0. \end{cases} \qquad (26)$$

$$HH'''(x, y) = \begin{cases} HH''(x, y) + 4, & \text{if } h_2''(j) > 0, \\ HH''(x, y) - 4, & \text{if } h_2''(j) < 0. \end{cases} \qquad (27)$$

$$HH2'''(x, y) = \begin{cases} HH2''(x, y) + 4, & \text{if } h_3''(j) > 0, \\ HH2''(x, y) - 4, & \text{if } h_3''(j) < 0. \end{cases} \qquad (28)$$

**Embedding Data into *LH*:**
Step 7: Using the sub-band *HL* with the hidden message as the reference, a fourth difference histogram is created with the original sub-band *LH*. The secret message is then hidden into this difference histogram according to steps 1 to 6; as a result, $LH'''(x, y)$ will be created. This completes the embedding process.

**Obtaining Marked Image:**
Step 8: Reconstruct the marked image by utilizing the inverse of the two-level HDWT four-band sub-band coding on the sub-bands. We first reconstruct the *LL* sub-band and then reconstruct the marked image $\psi$. This procedure can be formulated as follows:

$$LL = IDWT(S_{LL2}, S_{LH2}, S_{HL2}, S_{HH2}'''), \qquad (29)$$

$$\psi = IDWT(S_{LL}, S_{LH}''', S_{HL}''', S_{HH}'''). \qquad (30)$$

**Preventing overlap and over/underflow**
A flag-bit is used to indicate whether the bin in the different histograms is overlap or not. The flag-bit is set to 1 if a bin value shifted by ±8 overlaps with one shifted by ±4. The flag-bit is set to 0 if no overlap occurs. These flag-bits and the values of T and f in the bitmap will ultimately all be compressed with an efficient compression tool based on the LZMA algorithm. The compressed result is then hidden into the reserved non-overlapping bins. According to experimental results, this process results in less than 2.74% overhead in full embedding capacity for many different types of images.

Though the generated pixel values in the marked image may be outside the allowable range, the method in [12][18] could be used to deal with the problem. The interval range is dynamically selected with the image characteristics that it will reduce distortions to minimum.

### 3.3. Data extracting and reversing algorithm

Before extracting the hidden message, receiver needs to verify whether or not the marked image $\psi$ has been modified. If there is more than one occurrence at $h_{\psi i}(j) = -1$, we can conclude that the marked image $\psi$ has been tampered with. The proposed scheme then stops the following extraction steps immediately. The extraction and recovery process is schematized in Figure 4.

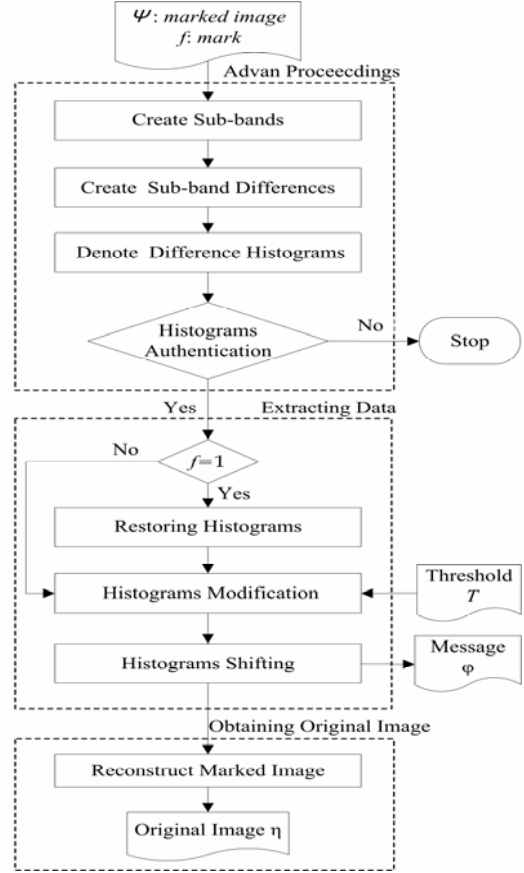

Figure 4. Flowchart of extraction and recovery process

The detailed extraction and recovery process includes the following steps:

**Extracting_process**$(\psi, T, f)$
Input: $\psi$, the marked image; $T$, the threshold; $f$, the mark, according to which the embedding status is determined.
Output: $\phi$, the secret message; $\eta$, the original image.

**Advan Proceedings:**
Step 1: Create sub-bands by performing two-level HDWT four-band coding on the marked image $\psi$. Five of the sub-bands to be utilized are denoted by $\psi_{LH}(x,y)$, $\psi_{HL}(x,y)$, $\psi_{HL}(x,y)$, $\psi_{LH2}(x,y)$ and $\psi_{HH2}(x,y)$, where $(x,y)$ indicates the coordinate of the coefficients in each sub-band.

**Recovering *LL* Sub-band**
Step 2: Create sub-band difference $D_{\psi 1}(x,y)$ between the reference sub-band $\psi_{LH2}$ and the destination sub-band $\psi_{HH2}$ according to the following formulas:

$$D_{\psi 1}(x, y) = \psi_{LH2}(x, y) - \psi_{HH2}(x, y). \quad (31)$$

Step 3: Denote the histogram of $D_{\psi 1}(x, y)$ as $h_{\psi 1}(j)$, where $(j)$ indicates the value of each bin.

Step 4: Check the distribution of the histogram $h_{\psi 1}(j)$. If there is more than one occurrence at $h_{\psi 1}(j) = -1$, the subsequent steps will be stopped immediately.

Step 5: Check the embedding status. Once the value of *f* is equal to 1, it can be concluded that the sub-band is completely filled with hidden message bits. Subsequently, first restore the original difference histogram. The bins greater than or equal to zero will be shifted to the right by 4 and those less than zero to the left by 4. The restored $h'_{\psi 1}(j)$ can be calculated as follows:

$$h'_{\psi 1}(j) = \begin{cases} h_{\psi 1}(j) + 4, & \text{if } h_{\psi 1}(j) \geq 0, \\ h_{\psi 1}(j) - 4, & \text{if } h_{\psi 1}(j) < 0. \end{cases} \quad (32)$$

These can also be obtained by the following formulas:

$$D'_{\psi 1}(x, y) = \psi_{LH2}(x, y) - \psi'_{HH2}(x, y), \quad (33)$$

where $\psi'_{HH2}(x, y)$ can be expressed as:

$$\psi'_{HH2}(x, y) = \begin{cases} \psi_{HH2}(x, y) - 4, & \text{if } h_{\psi 1}(j) \geq 0, \\ \psi_{HH2}(x, y) + 4, & \text{if } h_{\psi 1}(j) < 0. \end{cases} \quad (34)$$

**Extracting Data:**
Step 6: Extract the hidden message $\phi(n)$, where n denotes the index of a message bit, by shifting $h'_{\psi 1}(j)$ with reference to the bitmap and inverting the embedding process. First, the iteration index $\ell$ is set to 0. Once a $h'_{\psi 1}(j)$ with a value of $\pm(\ell + 4)$ is encountered, a binary bit "1" is retrieved. On the other hand, a binary bit "0" is retrieved if $h'_{\psi 1}(j)$ has a value of $\pm(\ell + 8)$. This procedure is repeated until there are no $h'_{\psi 1}(j)$ values of $\pm(\ell + 4)$ or $\pm(\ell + 8)$. Subsequently, $\ell$ is increased by 1. The same procedures as described above are repeated until $\ell$ reaches T+1. The retrieving rule is as follows:

$$h''_{\psi 1} = \begin{cases} h'_{\psi 1}(j) - 8, & \text{if } h'_{\psi 1}(j) = 8, \\ h'_{\psi 1}(j) - 4, & \text{if } h'_{\psi 1}(j) = 4, \end{cases} \quad (35)$$

for $\ell = 0$.

$$h''_{\psi 1}(j) = \begin{cases} h'_{\psi 1}(j) - 8, & \text{if } h'_{\psi 1}(j) = \ell + 8, \\ h'_{\psi 1}(j) - 4, & \text{if } h'_{\psi 1}(j) = \ell + 4, \\ h'_{\psi 1}(j) + 8, & \text{if } h'_{\psi 1}(j) = -(\ell + 8), \\ h'_{\psi 1}(j) + 4, & \text{if } h'_{\psi 1}(j) = -(\ell + 4), \end{cases} \quad (36)$$

for $1 \leq \ell \leq T$.

Step 7: At the same time, the modified sub-band difference $D'_{\psi 1}(x, y)$ is also scanned and modified. The extracting operation can be expressed as the following formula:

$$\phi(n) = \begin{cases} 0, & \text{if } D'_{\psi 1}(x, y) = \pm(\ell+8), \\ 1, & \text{if } D'_{\psi 1}(x, y) = \pm(\ell+4). \end{cases} \quad (37)$$

This procedure is executed until $\ell = T+1$.

Step 8: Remove the hidden message bits $\phi(n) \in \{0,1\}$ from the sub-band difference. The removing rule is given by

$$D''_{\psi 1}(x, y) = \psi_{LH2}(x, y) - \psi''_{HH2}(x, y), \quad (38)$$

where $\psi''_{HH2}(x, y)$ can be expressed as

$$\psi''_{HH2}(x, y) = \begin{cases} \psi'_{HH2}(x,y)+8, & \text{if } h'_{\psi 1}(j)=8, \; \varphi(n)=0, \\ \psi'_{HH2}(x,y)+4, & \text{if } h'_{\psi 1}(j)=4, \; \varphi(n)=1, \end{cases} \quad (39)$$

for $\ell = 0$, and

$$\psi''_{HH2}(x, y) = \begin{cases} \psi'_{HH2}(x,y)+8 & \text{if } h'_{\psi 1}(j)= \ell+8, \; \varphi(n)=0, \\ \psi'_{HH2}(x,y)+4 & \text{if } h'_{\psi 1}(j)= \ell+4, \; \varphi(n)=1, \\ \psi'_{HH2}(x,y)-8 & \text{if } h'_{\psi 1}(j)= -(\ell+8), \; \varphi(n)=0, \\ \psi'_{HH2}(x,y)-4 & \text{if } h'_{\psi 1}(j)= -(\ell+4), \; \varphi(n)=1, \end{cases} \quad (40)$$

for $1 \le \ell \le T$.

Step 9: Restore the original histogram. The original histogram $h'''_{\psi 1}$ can be calculated according to the following rules:

$$h'''_{\psi 1}(j) = \begin{cases} h''_{\psi 1}(j)-8 & \text{if } h''_{\psi 1}(j) \ge T+1, \\ h''_{\psi 1}(j)+8 & \text{if } h''_{\psi 1}(j) \le -(T+1). \end{cases} \quad (41)$$

for $\ell \le L$.

The restoration of the histogram as described above results in changes in the coefficients. This can be described by the following formula:

$$D'''_{\psi 1}(x, y) = \psi_{LH2}(x, y) - \psi'''_{HH2}(x, y). \quad (42)$$

Here, $\psi'''_{HH2}(x, y)$ can be expressed as:

$$\psi'''_{HH2}(x, y) = \begin{cases} \psi''_{HH2}(x,y)+8, & \text{if } h''_{\psi 1}(j) \ge T+1, \\ \psi''_{HH2}(x,y)-8, & \text{if } h''_{\psi 1}(j) \le -(T+1). \end{cases} \quad (43)$$

**Obtaining the *LL* Sub-band:**

Step 10: Recover the original Sub-band LL through the inverse operation of the HDWT algorithm with $\psi_{LL2}(x,y)$, $\psi_{LH2}(x,y)$, $\psi_{HL}(x,y)$, and $\psi_{HH2}^{m}(x,y)$. This procedure can be formulated as follows:

$$LL = IDWT(\psi_{LL2}, \psi_{LH2}, \psi_{HL2}, \psi_{HH2}^{m}). \qquad (44)$$

**Recovering *LH*, *HL* and *HH* Sub-bands:**

Step 11: After recovering the *LL* sub-band, steps 1 to 9 are repeated to recover the *LH*, *HL* and *HH* sub-bands as $\psi_{LH}^{m}(x,y)$, $\psi_{HL}^{m}(x,y)$ and $\psi_{HH}^{m}(x,y)$.

**Obtaining Original Image:**

Step 12: Recover the original image $\eta$ through the inverse operation of the HDWT algorithm with $\psi_{LH}^{m}(x,y)$, $\psi_{HL}^{m}(x,y)$, $\psi_{HH}^{m}(x,y)$ and $\psi_{LL}(x,y)$. This procedure can be formulated as follows:

$$\eta = IDWT(\psi_{LL}, \psi_{LH}^{m}, \psi_{HL}^{m}, \psi_{HH}^{m}). \qquad (45)$$

The above steps complete the extraction and recovering procedure.

## 4. EXPERIMENTAL RESULTS AND COMPARISON

In this section, a set of experiments are conducted to evaluate the embedding performance of our scheme. For these experiments, we used many different types of images, including some commonly used ones and two medical images (Figure 5). The message bits to be embedded in our experiments are randomly generated by a pseudo-random binary generator. The threshold ranges from 0 to 100.

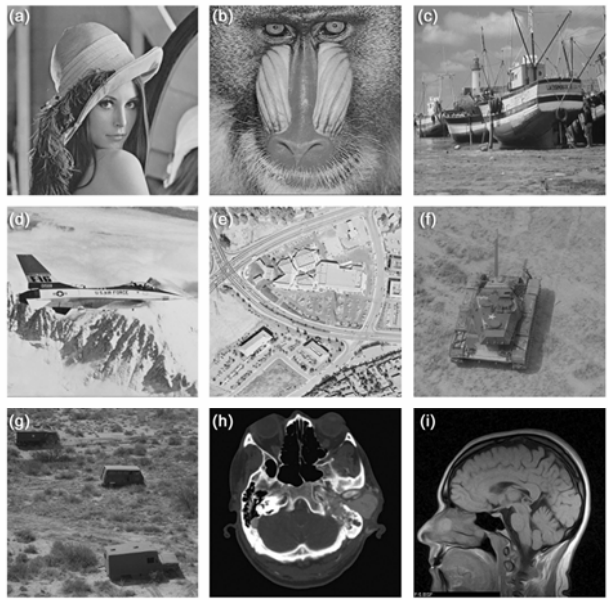

Figure 5. 8-bit 512×512 images(a)Lena, (b)Baboon, (c)Boat, (d)Airplane, (e)Aerial, (f)Tank, (g)Trucks, (h)Medical image1, (i) Medical image2

**Capacity versus Threshold**

Figure 6 depicts relationship between the capacity in bpp and threshold. The embedding rate almost reaches 0.78 bpp at threshold 100 for most the test images. As expected the capacity is nearly proportional to the threshold at the beginning and saturates when the threshold is sufficiently high.

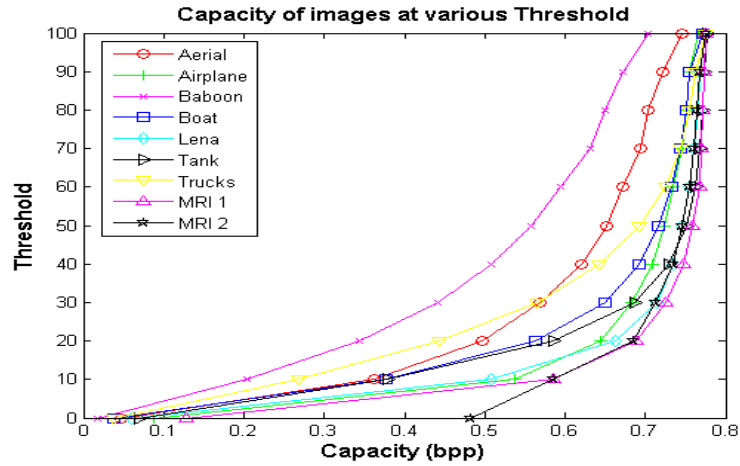

Figure 6. Embedding capacity at various thresholds

**Visual Quality versus Threshold**
Figure 7 illustrates the visual quality in PSNR of the marked image versus threshold varying from 0 to 100 for test images on the premise that maximal bits are embedded. The experimental result indicates that the PSNR rises as the threshold increases and this is not the case for the previous works. The marked image can achieve 43.3 dB at the threshold 0, and above 46 dB at the threshold of 100 for most test images. It's noteworthy that the larger threshold T can contribute less variation in the histogram of wavelet transform. Since the middle/high-frequency sub-bands incorporate less energy, the test images with larger variance between middle and high-wavelet coefficients such as "MRI 1" and "MRI 2" can achieve higher visual quality than "Baboon" at the threshold 0.

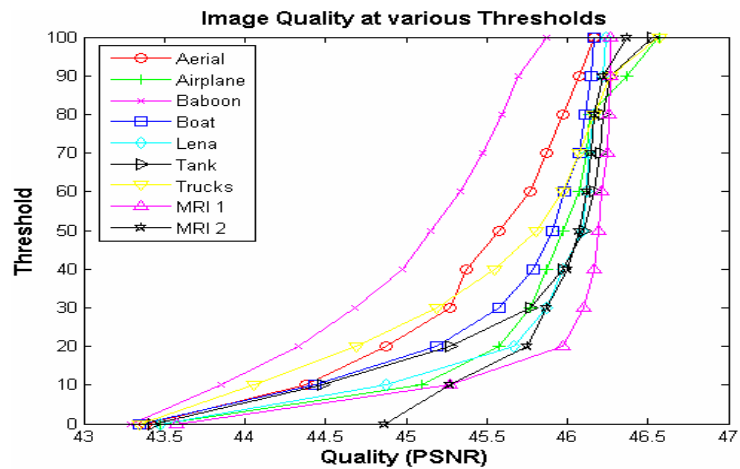

Figure 7. PSNR at various thresholds

**Comparison of visual quality with other schemes**
In terms of actual embedding capacity and visual quality, the proposed scheme was also compared with the DE scheme [14], G-LSB scheme [5], Kim et al.'s scheme [9], and Ni et al.'s scheme [6] for the "Lena", "Baboon", "Boat", and "Airplane" images as shown in Figure 8. The embedding capacity is the amount of embedded bits with overhead subtracted. It can be observed that our scheme is stable and achieves highest PSNR at the same bpp.

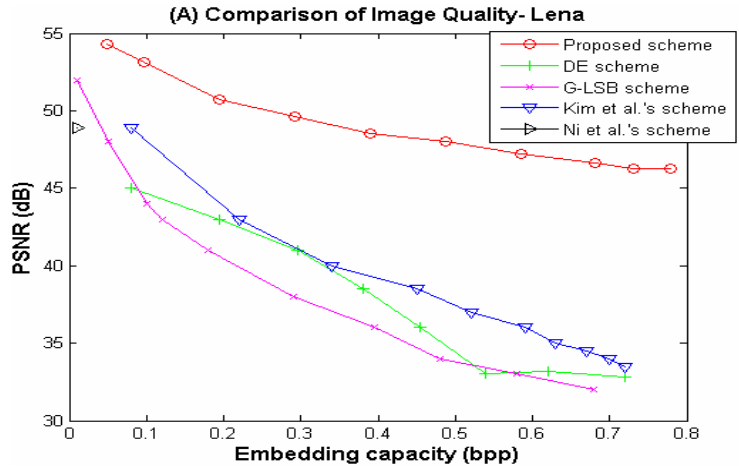

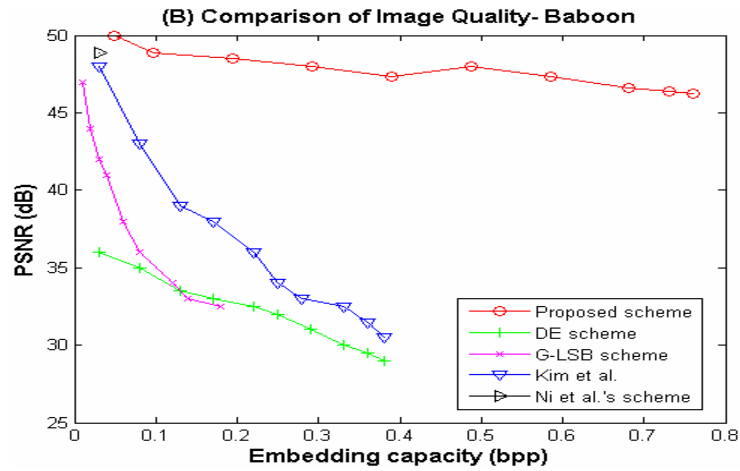

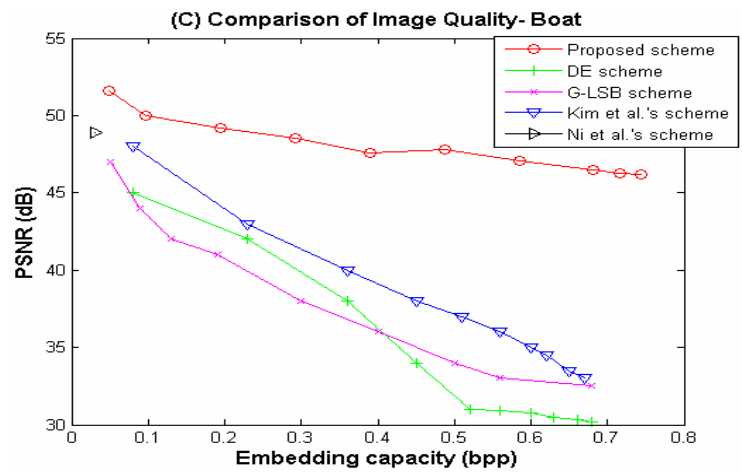

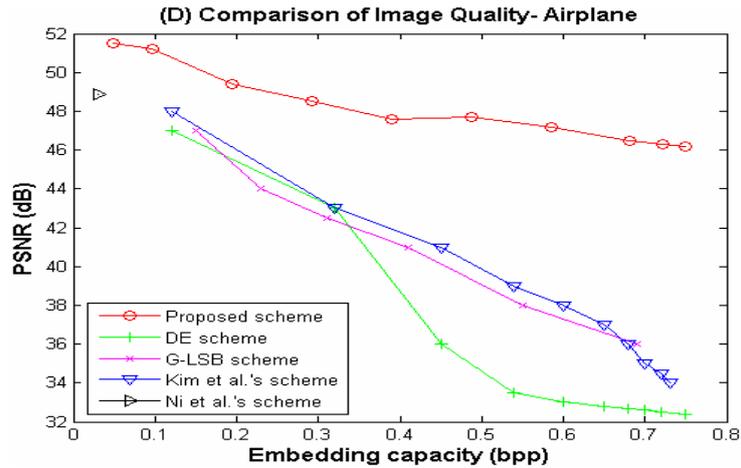

Figure 8. Comparison of embedding capacity in bpp versus distortion with existing reversible schemes: DE scheme, G-LSB scheme, Kim et al.'s scheme, and Ni et al's scheme: (A) Baboon; (B) Lena; (C) Boat; (D) Airplane

## 5. CONCLUSIONS

In this paper, we propose a simple and high-performance reversible data hiding scheme, which utilizes the large variance of wavelet coefficients and ingenious histogram shifting rules to avoid the decimal problem in pixel values during recovery process. Our scheme, compared with those reported previously, can obtain better visual quality of the marked images given the same payload. The main reason is that the visual quality of our scheme does not decay with increasing threshold as in the other schemes because the larger threshold can contribute less variation to the sub-band difference histograms. In addition, our scheme provides the greatest embedding capacity under nearly equivalent visual quality, as the particularities of large wavelet coefficient variance and minor changes in the wavelet coefficients following the embedding process are utilized. It may be of interest for future research that the threshold predictions, multi-round schemes, and fast algorithms will be explored to meet real-time application requirements.

**Authors**

**Xu-Ren Luo** received the B.S. degree in Computer Science and the M.S. degree in Electrical and Electronic Engineering, both from Chung Cheng Institute of Technology (CCIT), National Defense University, Taiwan, R.O.C., in 1997 and 2004, respectively. He is currently pursuing the Ph.D. degree in the area of information assurance at the Electrical and Electronic Engineering Department, CCIT. His research interests are focused on information security, data hiding, multimedia security, and image processing.

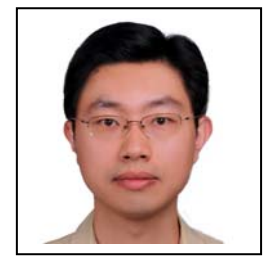

**Chen-Hui Jerry Lin** received the M.S. degree from the Department of Electrical Engineering, University of Missouri at Rolla in 1986. He then began his career of teaching computer programming at several colleges and universities of technology. He received the Ph.D. degree in electrical engineering from National Taiwan University in 1998. He has joined the faculty of National Defense University as an associate professor since 2002. His research interests include signal processing and computer programming.

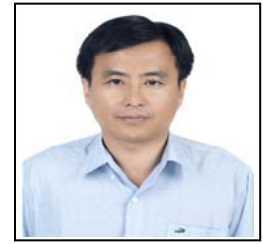

**Te-Lung Yin** received the B.S., M.S. and Ph.D. degree from the Chung Cheng Institute of Technology, National Defense University, Taiwan, R.O.C., in 1989, 1995 and 2001, respectively, in electrical engineering. He now is an associate professor of the Department of Computer Science and Information Engineering at the China University of Technology, Taiwan, R. O. C.. Dr. Yin is a member of the IEICE society and his research interests include data hiding, cryptography and image processing.

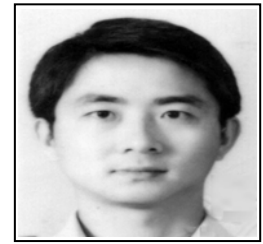